\documentclass[a4paper,10pt,twocolumn]{article} 
\usepackage{graphicx}
\usepackage{subfigure}
\usepackage{amsmath}
\usepackage{amssymb}
\usepackage{wrapfig}
\usepackage{color}
\usepackage[english]{babel}
\usepackage[warning,math]{easyeqn}
\usepackage[thinlines]{easybmat}
\begin{document}

\title{A simple Hubble-like law in lieu of dark energy}

\author{Yves-Henri Sanejouand\footnote{yves-henri.sanejouand@univ-nantes.fr}\\ \\
        UMR 6286 of CNRS,\\
        Facult\'e des Sciences et des Techniques,\\
        Nantes, France.} 
\maketitle


\section*{Abstract}
Within the frame of the $\Lambda$ cold dark matter paradigm, a dark energy component of unknown origin is expected to represent nearly 70\% of the energy of the Universe. Herein, a non-standard form of the Hubble law is advocated, with the aim of providing safe grounds on which alternative cosmologies could be developed. Noteworthy, it yields an age-redshift relationship which is consistent with available data. 
Together with a straightforward analysis of gamma-ray burst counts, it further suggests that the observable Universe has been euclidean and static over the last 12 Gyr. Although a non-standard distance-duality relation is then required for interpreting luminosity distances, the magnitude-redshift relationship obtained
is compatible with type Ia supernovae data.


\section{Introduction}
 Over the last twenty years, as a consequence of its numerous successes, 
 the $\Lambda$ cold dark matter ($\Lambda$CDM)
paradigm  has reached the status of a "concordance" cosmology~\cite{Ostriker:95,Frieman:08}. 
However, several clouds are still obscuring the brilliance of this paradigm,
 one of the most notable being that it relies on a new kind
 of so-called "dark energy", of unknown origin but accounting for at least 
 68\%~\cite{Planck:14}, and up to 75\% of the energy of the Universe~\cite{Frieman:08,Bartelmann:10}.
 
  As long as this dominant component remains mysterious~\cite{Li:11}, alternative cosmologies
 need to be developed~\cite{Sahni:06,Buchert:08,Marmet:12}. 
The purpose of the present study is to provide safe grounds
 on which such cosmologies could be established. 
 
 \section{Main hypothesis}

 In the late 1920s, Edwin Hubble discovered a proportionality between $z_\lambda$,
 the redshift of nearby galaxies, and $D_{mes}$, their  distance estimates~\cite{Hubble:29}.      
 He wrote his law as follows: 
\begin{equation}
z_\lambda = \frac{H_0 D_{mes}}{c_0}
\label{eq:Hubble}
\end{equation}
where $H_0$ is the Hubble constant, $c_0$, the speed of light, with $z_\lambda$ being:
\begin{equation*}
z_\lambda = \frac{\lambda_{obs} - \lambda_0}{\lambda_0} 
\end{equation*}
where $\lambda_{obs}$ is the wavelength of the light received from the galaxy, while $\lambda_0$ is the wavelength measured for the same kind of source sitting on Earth.

In the late 1990s, using type Ia supernovae as standard candles, it was shown that, for large values of the distance, Hubble's law is not linear any more~\cite{Riess:98sh,Perlmutter:99}. Within the frame of $\Lambda$CDM, this deviation from linearity is in particular due to a non-zero, although very small~\cite{Sahni:00}, value of $\Lambda$, the cosmological constant.

Hereafter, it is instead assumed that, as proposed by Hubble, the law he
discovered is indeed linear. However, it is also posited that, when Hubble defined the redshift, in the nowadays standard way, he made the wrong choice. Specifically, herein, the physically relevant form of Hubble's law is assumed to be:
 \begin{equation}
z_\nu = H_0 \Delta t
\label{eq:Hubble3}
\end{equation}
where $\Delta t$ is the photon time-of-flight between the source and the observer, $H_0$ being an actual constant, and where $z_\nu$, the frequency-redshift, is: 
 \begin{equation*}
z_\nu = \frac{\nu_0 - \nu_{obs}}{\nu_0}
\end{equation*}
$\nu_{obs}$ being the frequency of the light received from a remote source. 
Note that with $D_{mes} = c_0 \Delta t$,
when $z_\lambda \ll 1$, eqn~\eqref{eq:Hubble3} can indeed be approximated by eqn~\eqref{eq:Hubble} since, by definition:
\begin{equation}
z_\nu = \frac{z_\lambda}{1 + z_\lambda}
\label{eq:znu}  
\end{equation}    

\section{Age-redshift relationship}

\begin{table}[tbp]
\caption{The two oldest objects presently known.
HD140283, an extremely metal-deficient subgiant, 
is the oldest star known in our neighbourhood. 
APM 08279+5255 is the oldest quasar known at $z_\lambda \approx 4$.
The age of APM 08279+5255 was obtained through the measure of the Fe/O abundance ratio~\cite{Komossa:02}.}            
\label{tbl:ages}      
\begin{center}
\begin{tabular}{c c c c c}     
\hline\hline               
Object & $z_\lambda$ & $z_\nu$ & Age & Ref. \\    
       &             &         & (Gyr) &    \\
\hline                        
 HD140283 & 0 & 0 & 14.5 & \cite{Bond:13} \\      
 APM 08279+5255 & 3.9 & 0.8 & 2.1 & \cite{Friaca:05} \\
\hline                                   
\end{tabular}
\end{center}
\end{table}

\subsection{The age of the oldest stars}

Considering the case of early-type stars or galaxies born $T_{old}$ Gyr ago, $T_{obs}$, their apparent age according to an Earth-based observer is:
\begin{equation*}
T_{obs} = T_{old} - \Delta t
\end{equation*}
that is, with eqn~\eqref{eq:Hubble3}:
\begin{equation}
T_{obs} = T_{old} - T_H z_\nu
\label{eq:age}
\end{equation}
or, with eqn~\eqref{eq:znu}:
\begin{equation}
T_{obs} = T_{old} - T_H \frac{z_\lambda}{1 + z_\lambda}
\label{eq:agez}
\end{equation}
where $T_H = \frac{1}{H_0}$ is the Hubble time. 
Table~\ref{tbl:ages} shows the age estimates of what may be the two oldest objects presently known, at their respective redshift. 

\begin{figure}[tbp]
\vskip 0.2 cm
\includegraphics[width=7.5 cm]{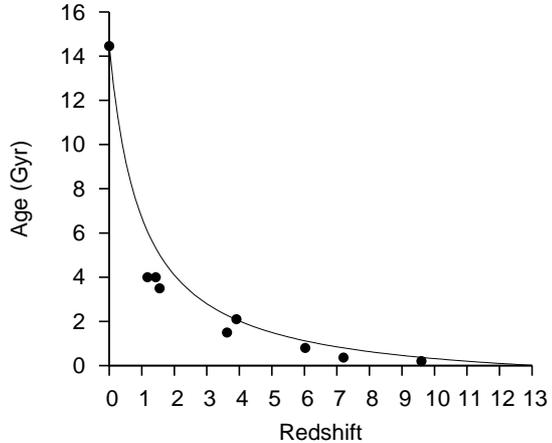}
\caption{Age-redshift (z$_\lambda$) relationship. Filled circles: ages of nine early-type stars, galaxies and quasars (see text). Plain line: upper bound expected with $T_{old}$=14.5 and T$_H$=15.6 Gyr, according to the Hubble-like law advocated herein.}
\label{Fig:agez}
\end{figure}

Being in our neighbourhood, HD140283 can provide a fair estimate for $T_{old}$ while, according to eqn~\eqref{eq:age} and under the hypothesis that APM 08279+5255 could be, nowadays, as old as HD140283, the apparent age ($T_{obs} =$ 2.1 Gyr) and frequency-redshift ($z_\nu =$ 0.8) of the former allow to determine the Hubble time, namely:
\begin{equation*}
T_H = \frac{T_{old} - T_{obs}}{z_\nu} = 15.6~Gyr
\end{equation*}
that is, a value consistent with recent estimates~\cite{Planck:14,Sandage:06,Melia:13}. 
Note that the above analysis is expected to yield an accurate value
for $T_H$ if HD140283 and APM 08279+5255 have, nowadays, the same age.
Indeed, eqn~\eqref{eq:agez} provides an upper-bound for the age of early-type stars at any redshift. 

For instance, it has been claimed that two galaxies found at $z_\lambda =$ 6 and $z_\lambda =$ 9.6 could be 0.8~\cite{Fiedler:11} and 0.2~\cite{Wel:12sh} Gyr old, respectively. According to eqn~\eqref{eq:agez}, and as illustrated in Fig.\ref{Fig:agez}, the corresponding upper-bounds at these redshifts are indeed higher, namely, 1.2 and 0.3 Gyr, respectively.
Fig.\ref{Fig:agez} also shows that eqn~\eqref{eq:agez} yields upper bounds that are over current estimates for the ages of 3C65~\cite{Stockton:95}, LBDS 53W069~\cite{Dunlop:98}, LBDS 53W091~\cite{Dunlop:96}, QSO B1422+231~\cite{Yoshii:98} and GNS-zD1~\cite{Brammer:10}, at $z_\lambda$=1.175, 1.43, 1.55, 3.62 and 7.2, respectively.   

\subsection{A corollary}

As a consequence of eqn~\eqref{eq:age}, $T_{obs} > 0$ if:
\begin{equation}
z_\nu < \frac{T_{old}}{T_H}
\label{eq:zmax}
\end{equation}     
that is, with eqn~\eqref{eq:znu} and the above values for $T_{old}$ and $T_H$:
\begin{equation*}
z_\lambda < 13
\end{equation*}
So, according to eqn~\eqref{eq:agez}, and under the assumption that no object older than HD140283 or APM 08279+5255 can be observed nowadays from Earth, it should not be possible to observe any galaxy at a redshift larger than 13.
This is consistent with current knowledge, since the highest redshifts known so far are around 10~\cite{Wel:12sh,Bradley:11sh}.  
Note that this upper limit would drift if objects older than HD140283 or APM 08279+5255
are discovered. 

\subsection{Another prediction}

\begin{figure}[tbp]
  \includegraphics[width=7.5 cm]{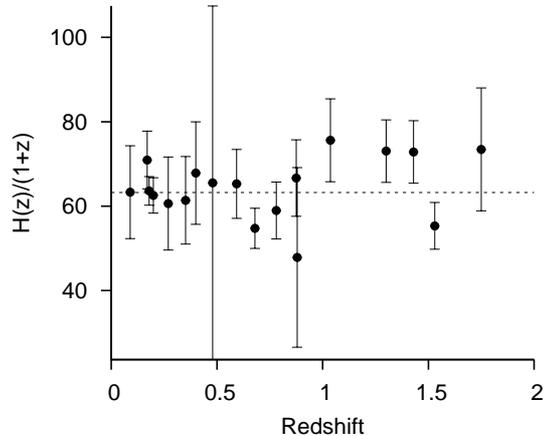}
   \caption[]{ $\frac{H(z_\lambda)}{1+z_\lambda}$ as a function of redshift ($z_\lambda$). The dashed line corresponds to $H_0 =$ 63 km.s$^{-1}$.Mpc$^{-1}$ ($T_H =$ 15.6 Gyr). $H(z_\lambda)$ data come from~\cite{Jimenez:12}.}
\label{Fig:hz}
\end{figure}

On the other hand, as a consequence of eqn~\eqref{eq:Hubble3}:
\begin{equation*}
\frac{\partial z_\nu}{\partial(\Delta t)} = H_0 
\end{equation*}
With $\Delta t=t_0 - t$, taking eqn~\eqref{eq:znu} into account yields: 
\begin{equation}
\frac{\partial z_\lambda}{\partial t} = - H_0 (1 + z_\lambda)^ 2
\label{eq:esp}
\end{equation}
where $t_0$ and $t$ are the observer and cosmic times, respectively. 
Measures of $\frac{\partial z_\lambda}{\partial t}$, obtained through studies of the age of passively evolving galaxies, are usually provided through $H(z_\lambda)$~\cite{Stern:10,Jimenez:12}, which is defined as follows~\cite{Jimenez:02}:
\begin{equation*}
H(z_\lambda)= -\frac{1}{1 + z_\lambda}\frac{\partial z_\lambda}{\partial t}
\end{equation*}
that is, with eqn~\eqref{eq:esp}:
\begin{equation}
\frac{H(z_\lambda)}{ 1 + z_\lambda } = H_0 
\label{eq:hz}
\end{equation}
The corresponding relationship expected within the frame of $\Lambda$CDM is not that simple~\cite{Jimenez:02,Ratra:13}. As a matter of fact, it has been claimed that eqn~\eqref{eq:hz}, which is also a prediction of the $R_h = c_0 t$ cosmology~\cite{Melia:07,Melia:12}, is ruled out by the data~\cite{Bilicki:12}. However,
backed by standard statistical analysis ($\chi^2$ = 13.4, p-value = 0.71),
and in spite of large error bars (see Fig.\ref{Fig:hz}),
a recent in-depth study shows that, on the contrary, when compared to $\Lambda$CDM, eqn~\eqref{eq:hz} is favoured by model selection criteria~\cite{Melia:13}.
 
\section{Distance versus redshift}

\subsection{Gamma-ray bursts}

Without any explicit cosmology, going further requires additional hypotheses.

So, in order to get insights about the relationship between photon time-of-flight and distance measurements, let us turn to cumulative object counts
and assume that:
\begin{equation}
n(z)=\beta D_c^d
\label{eq:count}
\end{equation}
where $n(z)$ is the number of objects with a redshift lower than $z_\lambda$, $\beta$,
a constant,
$d$ being the effective dimension of space when $D_c$ is the light-travel distance, namely:
\begin{equation}
D_c = c_0 \Delta t
\label{eq:dc}
\end{equation}
With eqn~\eqref{eq:Hubble3}, \eqref{eq:znu} and \eqref{eq:dc}, eqn~\eqref{eq:count} becomes:
\begin{equation}
n(z)=\beta D_H^d (\frac{z_\lambda}{1+z_\lambda})^d
\label{eq:countz}
\end{equation}
where $D_H = c_0 T_H$ is the Hubble length.

In order to measure $d$, it is necessary to consider a category of objects whose redshift is known over a wide range, with few selection biases. In this respect, noteworthy because they are highly energetic, sources of gamma-ray bursts (GRB) are attractive candidates. 

\begin{figure}[tbp]
\vskip 0.1cm
\hskip -1 cm
  \includegraphics[width=8.5 cm]{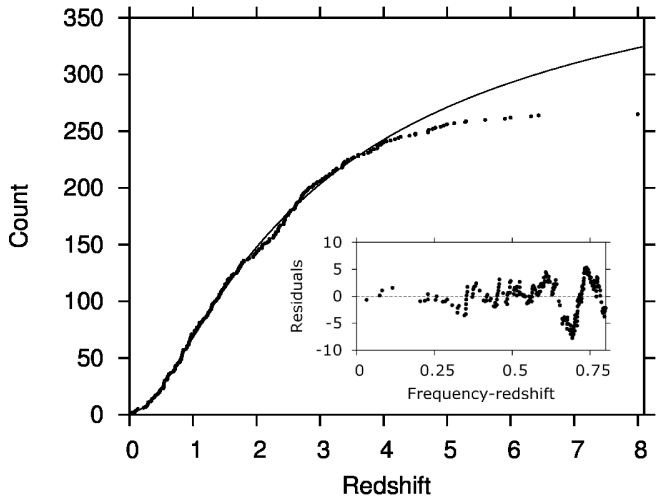}
   \caption[]{ Cumulative count of long gamma-ray bursts ($t_{90} > 0.8$ s), as a function of redshift ($z_\lambda$). Plain line: least-square fit of the data, performed for GRB sources with $z_\lambda < 3.5$. Inset: residuals (observed$-$expected), as a function of frequency-redshift ($z_\nu$).}
\label{Fig:grb}
\end{figure}

Thanks to the \textit{Swift} mission launched ten years ago~\cite{Swift,Derek:09}, the redshifts of 265 GRB sources have nowadays\footnote{http://swift.gsfc.nasa.gov, 2014, July 5$^{th}$.} been determined with fair accuracy. Moreover, in the \textit{Swift} sample, most GRB with a duration over 0.8 s are expected to have same physical origin~\cite{Sari:13}. 
Fig.\ref{Fig:grb} shows the cumulative count for the corresponding subset of 254 GRB sources, which is expected to be rather homogeneous. 

With eqn~\eqref{eq:countz},  
a least-square fit of these data yields:
\begin{equation*}
d = 2.96 \pm 0.03
\end{equation*}
Such a result strongly suggests that GRB sources are randomly distributed in an observable Universe that is both euclidean and static.
Indeed, in such a case, $d=3$ and 
$\beta = \frac{4}{3} \pi \rho_{grb}$,
where $\rho_{grb}$ is the average density of observable GRB sources.  
Note that this result relies on the hypothesis that Swift data
represent a fair sample of the GRB sources, up to $z_\lambda \approx 4$. 

Indeed,
above this value,
eqn~\eqref{eq:countz} starts to overestimate the observed GRB counts.
Such an over-prediction could mean that GRB happened to be less frequent more than 12 Gyr ago (with T$_H$ = 15.6 Gyr). It could also mean that, due to the sensitivity limits of \textit{Swift} and follow-up telescopes that were used to measure the redshifts, a significant fraction of the GRB with $z_\lambda > 4$ were missed.

Oscillations around values predicted by eqn~\eqref{eq:countz} may also prove meaningful (see the inset of Fig.~\ref{Fig:grb}) since, near $z_\nu \approx 0.7$ ($z_\lambda \approx 2.3$), the residuals exhibit a large-scale fluctuation of the GRB density, whose size ($\approx$ 400 Mpc; $\delta z_\nu \approx 0.1$) is of the order of the size of the largest voids known in our neighbourhood~\cite{Shandarin:82,Void}. 

\subsection{Angular diameter distance}

If, as suggested by the above results, the Universe is both euclidean and static, $D_A$, the angular diameter distance, is so that:
\begin{equation}
D_A = D_c 
\label{eq:da}
\end{equation}
Taking into account eqn~\eqref{eq:Hubble3}, \eqref{eq:znu} and \eqref{eq:dc} yields:
\begin{equation}
D_A = D_H \frac{z_\lambda}{1 + z_\lambda}
\label{eq:dofz}
\end{equation}
As a consequence, $\theta$, the angular size:
\begin{equation*}
\theta = \frac{s}{D_A}
\end{equation*}
$s$ being the actual size of the considered standard rod, becomes:
\begin{equation}
\theta = \frac{s}{D_H} ( 1 + \frac{1}{z_\lambda} )  
\label{eq:theta}  
\end{equation}
Indeed, although this may not be the case for ultra-compact~\cite{Jackson:06} or double-lobed radio sources~\cite{Buchalter:98}, it has been claimed that,
over a wide range of redshifts (up to $z_\lambda =$ 3.2), 
the average angular size of galaxies is approximately proportional to $z_\lambda^{-1}$~\cite{Lopez:10,Scarpa:14}.
Note that, assuming the standard cosmological model as correct, this fact can be explained only if the average linear size of galaxies with same luminosity is six times smaller at $z_\lambda =$ 3.2 than at $z_\lambda =$ 0~\cite{Lopez:10}.

\begin{figure}[tbp]
\hskip -0.8cm 
\includegraphics[width=8.5 cm]{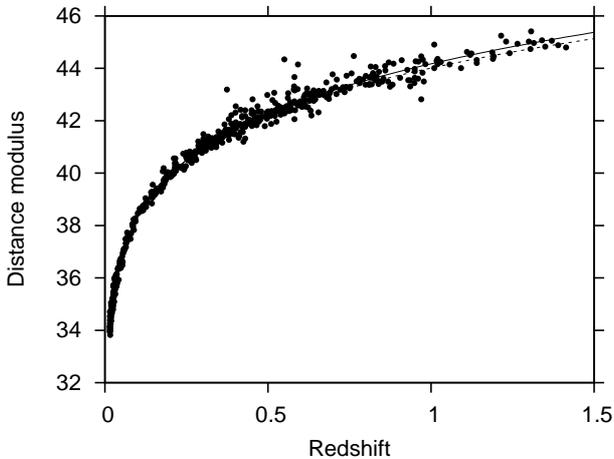}
\caption[]{The distance modulus of supernovae Ia, as a function of redshift (z$_\lambda$). Filled circles: the 580 cases of the Union 2.1 compilation (error bars not shown). Plain-line: least square fit of these data ($n=$ 1.65 $\pm$ 0.02, $\mu_0=$ 18.21 $\pm$ 0.02). Dashed line: the $n=\frac{3}{2}$ case ($\mu_0=$ 18.30 $\pm$ 0.01).}
\label{Fig:Sn}
\end{figure}

\subsection{Luminosity distance}

The luminosity distance, $D_L$, is related to the angular distance through the distance-duality relation, that is:
\begin{equation}
D_L = D_A ( 1 + z_\lambda )^n
\label{eq:dual}
\end{equation}

Together with eqn~\eqref{eq:Hubble3} and \eqref{eq:znu}, \eqref{eq:dual} allows to write $\mu$, the distance modulus:
\begin{equation*}
\mu = 5 \log_{10} ( D_L ) + 25
\end{equation*}
as follows:
\begin{equation}
\mu = 5 \log_{10} (z_\lambda ( 1 + z_\lambda )^{n-1}) + \mu_0 + 25
\label{eq:dmod}
\end{equation}
where $\mu_0 = 5 \log_{10} (D_H )$. 

Nowadays, distance moduli have been measured for hundreds of supernovae of type Ia (SNIa)~\cite{Kowalski:08}. As shown in Fig.\ref{Fig:Sn} for the 580 cases of the Union 2.1 compilation~\cite{Union2:12sh}, an accurate least-square fit of the data ($\chi^2 =$ 571, p-value = 0.57)\footnote{The error estimates on the distance modulus measurements were used for the $\chi^2$ calculation. This amounts to assume that SNIa are \textit{perfect} standard candles.} can be obtained with eqn~\eqref{eq:dmod}, which yields:
\begin{equation*}
n = 1.65 \pm 0.02
\end{equation*}
Note that when another type Ia supernova dataset is considered, namely, the 397 cases of the Constitution compilation~\cite{Hicken:09sh}, the value found for $n$ is similar ($n=1.63 \pm 0.03$). 

Let us emphasize that, within the frame of $\Lambda$CDM, like in most cosmologies based on a metric theory of gravity, $n=2$~\cite{Uzan:04,Holanda:10}. 
Indeed, in this context, $n$ can \textit{not} be lower than two while,
in the context of a static Universe, the most likely values are either $n=0.5$, as a consequence 
of the energy loss of the photons during their flight towards the observer, or $n=1$, 
if time dilation of SNIa lightcurves is also taken into account~\cite{Perlmutter:96,Riess:96,Blondin:08}. 
However, if for instance the number of photons is not conserved during their travel, $n$ can be larger than that~\cite{Kunz:04}. 

\begin{figure}[tbp]
\hskip -0.5cm
\includegraphics[width=8.5 cm]{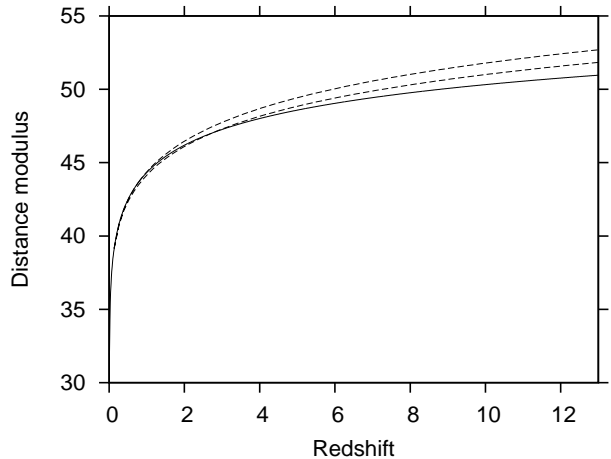}
\caption[]{Distance modulus, as a function of redshift (z$_\lambda$). Plain line: $\Lambda$CDM, with $\Omega_m$ = 0.3 and $\Omega_\Lambda$ = 0.7; dashed lines:  $n=$ 1.65 (above) and $n=\frac{3}{2}$ (below).}
\label{Fig:lcdm}
\end{figure}

\subsection{The distance-duality relation}
\label{sec:duality}

Deviation from the Etherington relation (eqn~\eqref{eq:dual}) has been quantified through the $\eta_0$ parameter, which can be defined as follows:
\begin{equation}
D_L = D_A ( 1 + z_\lambda )^{2+\eta_0}
\label{eq:eta}
\end{equation}
that is, for small values of $z_\lambda$:
\begin{equation*}
D_L \approx D_A ( 1 + z_\lambda )^2 ( 1 + \eta_0 z_\lambda )
\end{equation*}
By combining the Sunyaev-Zeldovich effect and X-ray surface brightness for two samples of galaxy clusters~\cite{Filippis:05,Bonamente:06}, together with type Ia supernovae data so as to end with a model-independent cosmological test, it was shown that $\eta_0 = -0.28 \pm 0.22$~\cite{Holanda:10}, when a sample of 25 galaxy clusters~\cite{Filippis:05} is analysed, and $\eta_0 = -0.42 \pm 0.11$~\cite{Holanda:10}, when a larger sample of 38 galaxy clusters~\cite{Bonamente:06} is considered. In the later case, when a redshift bias is accounted for, $\eta_0 = -0.23 \pm 0.11$ or $\eta_0 = -0.43 \pm 0.10$, depending upon which type Ia supernovae dataset is taken into account~\cite{Liang:13}.
More recently, using a sample of 91 galaxy clusters and four different methods, 
$\eta_0$ values were found to range between $\eta_0 = -0.08 \pm 0.10$, and $\eta_0 = -0.17 \pm 0.17$~\cite{Holanda:15}.

So, while metric theories of gravity like $\Lambda$CDM require $\eta_0 = 0$, observed values have been found to favor the negative side, up to 4$\sigma$ away from the $\Lambda$CDM prediction. On the other hand, all of them but one are within 2$\sigma$ of $\eta_0 = -0.35$, the value expected within the frame of the present study (with n=1.65).   

\subsection{Another difference with $\Lambda$CDM}

As shown in Fig.\ref{Fig:lcdm}, as far as distance moduli are concerned, the difference between values predicted with $\Lambda$CDM or eqn~\eqref{eq:dmod} becomes obvious for $z_\lambda > 2$, when $n=$ 1.65, or for $z_\lambda > 4$, when $n=\frac{3}{2}$. Although the fit of the supernovae data of the Union 2.1 compilation looks poor when $n=\frac{3}{2}$ ($\chi^2 =$ 652, p-value = 0.02), it follows the values predicted by $\Lambda$CDM over a wider range of redshifts (Fig.\ref{Fig:lcdm}). 

Note that $n = \frac{3}{2}$ is expected within the frame of a couple of alternative cosmologies~\cite{Marmet:12,Brynjolfsson:04}.

\section{Possible meanings}

Eqn~\eqref{eq:Hubble3} is so simple that, like the original Hubble law itself~\cite{Marmet:12}, it can be derived in many different ways, based on a variety of physical ground~\cite{Heymann:14}. Noteworthy, it is a straightforward consequence of the $R_h=c_0 t$ Universe~\cite{Melia:07,Melia:12}. 
 
As another example, let us assume that, for some yet unknown reason, a steady drift of atomic and molecular spectra takes place~\cite{Sumner:94,Ranada:08,Paturel:08} such that, for any frequency:
\begin{equation}
\nu \propto t   
\label{eq:drift}
\end{equation}
Furthermore, let us also assume that, during its flight between the source and the observer, the energy of the photon is conserved, \textit{i.e.}, that its frequency does not change. As a consequence, $\nu_{obs}$, the frequency of the photon received from a remote source is the frequency the photon had when it was emitted at $t = t_0 - \Delta t$:
\begin{equation*}
\nu_{obs} \propto t_0 - \Delta t
\end{equation*}
And since, according to eqn~\eqref{eq:drift}, $\nu_0 \propto t_0$:
\begin{equation}
\frac{\nu_{obs}}{\nu_0} = 1 - \frac{\Delta t}{t_0}
\label{eq:zclock}
\end{equation}
In other words, if $t_0 = T_h$, that is, if $T_H$ is assumed to be the time elapsed since photons started to be emitted with non-vanishing frequencies then eqn~\eqref{eq:Hubble3}, the Hubble-like law advocated in the present study,  is recovered.

Of course, such a drift of atomic spectra should show up in various physical domains, noteworthy as a consequence of a corresponding drift of atomic clocks. In particular, an apparent increase of lengths measured through the time it takes for electromagnetic waves to go from a place to another should be observed~\cite{Sanejouand:09}.      

\section{Conclusion}

A Hubble-like law, where the frequency-redshift is proportional to the photon time-of-flight, yields an age-redshift relationship which is consistent with available data. 
A straightforward analysis of gamma-ray burst counts further suggests that the observable Universe has been euclidean and static over the last 12 Gyr. 

Through a non-standard distance-duality relation, which is consistent with current knowledge, it also yields an alternative explanation for the luminosity distance data, alleviating the need for a dark energy component of unknown origin. 

Overall, the present study provides a frame, namely, a background that is euclidean and static, as previously advocated by others~\cite{Lopez:10,Scarpa:14,Hartnett:11}, as well as a set of relationships between redshifts and distances which could become useful anchors for the development of new cosmologies.      

\section*{Acknowledgements}

I thank Georges Paturel for fruitful discussions, Maciej Bilicki for his useful comment,
and referee B of \textit{EPL} for his careful reading of the manuscript, as well as for his constructive suggestions.


\end{document}